\documentclass[rnote]{aa}

\usepackage[varg]{txfonts}
\usepackage{longtable}
\usepackage{graphicx}
\usepackage{float}

\newcommand{\teff}{$T_{\rm eff}$}
\newcommand{\feh} {\mbox{\rm [Fe/H]}}
\newcommand{\logg}{$\log g$}

\newcommand{\sh}{$S_{\!\!\rm H}$}

\newcommand{\alp}{$\alpha$}
\newcommand{\alpfe}{[$\alpha$/Fe]}

\newcommand{\cui} {\ion{Cu}{i}}
\newcommand{\cuii} {\ion{Cu}{ii}}
\newcommand{\cu}[7]{\mbox{${#1}{#2}^{#3}\,^#4{#5}^{\rm #6}_{#7}$}}
\newcommand{\cufe} {\mbox{\rm [Cu/Fe]}}

\newcommand{\eps}[1]{\log{\varepsilon}_\sun {\rm #1}}

\newcommand{\optd}{$\log \tau_{\rm 5000}$}
\newcommand{\bi}{$b_i$}
\newcommand{\nil}{$n_i^{\rm LTE}$}
\newcommand{\nin}{$n_i^{\rm non-LTE}$}

\usepackage{amstext}

\begin{document}

\title{Non-LTE analysis of copper abundances for the two distinct halo populations in the solar neighborhood
\thanks{Based on observations made with the FIbre fed Echelle Spectrograph (FIES) at the Nordic Optical Telescope (NOT) on La Palma, and on data from the European Southern Observatory ESO/ST-ECF Science Archive Facility (programs 65.L-0507, 67.D-0086, 67.D-0439, 68.D-0094, 68.B-0475, 69.D-0679, 70.D-0474, 71.B-0529, 72.B-0585, 76.B-0133 and  77.B-0507)}
\fnmsep\thanks{Figs. 4$-$7 and Tables 1$-$2 are only available in electronic form at 
\tt{http://www.aanda.org}}
}

\author{H.L.~Yan \inst{1,2}
\and J.R.~Shi \inst{1}
\and P.E.~Nissen \inst{3}
\and G.~Zhao \inst{1}
}

\institute{Key Laboratory of Optical Astronomy, National Astronomical Observatories, Chinese Academy of Sciences, 20A Datun Road, Chaoyang District, Beijing 100012, China. \\
\email{sjr@nao.cas.cn}
\and University of Chinese Academy of Sciences, No.19A Yuquan Road, Beijing 100049, China.
\and Stellar Astrophysics Centre, Department of Physics and Astronomy, Aarhus University, 8000 Aarhus C, Denmark. \\
\email{pen@phys.au.dk}
}

\date{Received date / Accepted date}

\abstract
{Two distinct halo populations were found in the solar neighborhood by a series of works. They can be clearly separated by \alpfe\ and several other elemental abundance ratios including \cufe. Very recently, a non-local thermodynamic equilibrium (non-LTE) study revealed that relatively large departures exist between LTE and non-LTE results in copper abundance analysis. The study also showed that non-LTE effects of neutral copper vary with stellar parameters and thus affect the \cufe\ trend.}
{We aim to derive the copper abundances for the stars from the sample of \citet{Nis10} with both LTE and non-LTE calculations. Based on our results, we study the non-LTE effects of copper and investigate whether the high-\alp  \ population can still be distinguished from the low-\alp \ population in the non-LTE \cufe\ results.}
{Our differential abundance ratios are derived from the high-resolution spectra collected from VLT/UVES and NOT/FIES spectrographs. Applying the MAFAGS opacity sampling atmospheric models and spectrum synthesis method, we derive the non-LTE copper abundances based on the new atomic model with current atomic data obtained from both laboratory and theoretical calculations.}
{The copper abundances determined from non-LTE calculations are increased by $0.01$ to $0.2$ dex depending on the stellar parameters compared with the LTE results. The non-LTE [Cu/Fe] trend is much flatter than the LTE one in the metallicity range $-1.6<\feh<-0.8$. Taking non-LTE effects into consideration, the high-  \
and low-\alp \ stars still show distinguishable copper abundances, which appear even more clear in a diagram of non-LTE \cufe\ versus \feh.}
{The non-LTE effects are strong for copper, especially in metal-poor stars. Our results confirmed that there are two distinct halo populations in the solar neighborhood. The dichotomy in copper abundance is a peculiar feature of each population, suggesting that they formed in different environments and evolved obeying diverse scenarios.}


\titlerunning{Non-LTE copper abundances for the two distinct halo populations}
\authorrunning{Yan et al.}

\maketitle

\section{Introduction}
\label{sec1}
To understand how the Milky Way formed and evolved is one of the key and fundamental questions of modern astrophysics. The stellar populations and the substructures in the Galaxy are the most natural indicators of the processes that the Milky Way experienced. Much effort has been invested into revealing the composition of the stellar populations in the Galactic halo. The classic work of \cite{Egg62} implies that the stars in the Galactic halo belong to one population, which originated from the material that accumulated  during the collapse of the protogalaxy. This single-component halo was considered to be plausible for a long time, yet different scenarios were raised in the following decades. \cite{Sea78} found clues from globular clusters indicating that there are two halo populations that can be divided according to their horizontal branch morphology. Since then, evidence from different works continued to challenge the single-component halo model \citep[e.g.,][]{Zin93, Car96, Wil96, Kin07, Lee07, Mic08}. Recent data from large survey programs also prefer a more complex Galactic halo that consists of at least two populations \citep{Car07, Jof11, Haw15}. 

The variation of relative abundances as a function of  stellar \feh\ is a powerful tool for identifying different populations
because various species of elements may have been synthesized on different timescales and in various environments \citep{Tin79}. The \alp-elements (e.g., Mg, Si, Ca, and Ti) are very important. In contrast to iron, which is mainly synthesized in type Ia supernovae (SNe Ia) on a relatively long timescale, the dominate source of \alp-elements is the shorter-lived type II supernovae (SNe II), therefore the \alpfe\ can represent an ideal cosmos-clock to trace the history of chemical evolution. If the halo is composed of more than one component, it would be reasonable to expect that systematic differences in the stellar properties also exist for the halo stars in the solar neighborhood, where the \alp-element abundances can be measured precisely based on high-resolution spectra. Previous studies have found variations in \alpfe\ between the inner- and outer-halo stars, but it is unclear whether there is a dichotomy in the \alpfe\ pattern or not \citep[see discussion in][]{Nis11}. In a series of works, \cite{Nis10}, \cite{Nis11} (Hereafter cited as NS10 and NS11, respectively), \cite{Sch12} and \cite{Nis12}  detected two distinct halo populations with dual \alpfe\ distributions in the solar neighborhood. The two populations also show differences in other properties, including \cufe.  NS11 found that the high-\alp \ stars have a tight \cufe\ trend that is similar to that of the thick-disk population, while the low-\alp \ stars have a much looser \cufe\ trend and a systematic \cufe\ deficiency compared to the high-\alp \ stars, presenting a clear distinction in the \cufe\ versus \feh\ diagram. 

However, the abundance analysis in NS11 was based on LTE assumption. Previous works on non-LTE effects have demonstrated the importance of its effect, especially for metal-poor stars \citep[e.g.,][]{Zha00,Geh04,Ber08,Shi09}. Copper is thought to be one of the elements that may suffer strong non-LTE effects. Several studies have reported that the abundances derived from some of the \cui\ or \cuii\ lines are inconsistent with each other \citep{Bih04, Bon10, Roe12, Roe14}. Very recently, \cite{Yan15} proposed a non-LTE analysis for the copper abundance in late-type metal-poor stars, showing that the value of \cufe\ can be underestimated by up to 0.2 dex in LTE analysis. The underestimation is correlated with stellar parameters, especially with the metallicity. It is therefore interesting to investigate whether the separation in \cufe\ reported in NS11 still exists when non-LTE effects are taken into consideration.

To derive a reliable \cufe\ trend and reveal the non-LTE effect on distinguishing different populations, we reanalyzed the copper abundances for the sample stars that were used in NS10 and NS11. The results are given with both LTE and non-LTE calculations. In Sect. \ref{sec2} we describe how we obtained the LTE and non-LTE copper abundances, including the details of non-LTE calculations. Section \ref{sec3} describes the results and error analysis. The conclusions are presented in Sect. \ref{sec4}.

\section{Methods}
\label{sec2}

\subsection{Non-LTE calculations}
\label{sec2.1}
We have presented the details of copper atomic model in our previous work \citep{Shi14}. In general, our final atomic model consists of $96$ energy levels of \cui\ plus the ground state of \cuii. Laboratory data from NIST\footnote{\tt http://www.physics.nist.gov/} and \citet{Sug90} were used for the 51 levels with low-excitation energy, while for the rest of the levels with higher excitation energy, we applied the theoretical work of \citet{Liu14}, which is based on the $R-$matrix approach \citep{Bur71, Liu11}. Fine structures were considered to the level of \cu{5}{p}{}{2}{P}{o}{}. Probabilities for all the 1089 bound-bound transitions applied for \cui\ and the photoionization cross-sections were also calculated by \citet{Liu14}. In addition, the excitation and ionization caused by inelastic collisions with electrons and hydrogen atoms were taken into account based on a series of theoretical works \citep[see details in][]{Shi14}. We scaled the collisional rates with neutral hydrogen by a factor of \sh$=0.1$ to avoid the overestimation \citep{Bar11} from the Drawin fomula \citep{Dra68, Dra69} following \citet{Shi14}. A revised version of the DETAIL program \citep{But85} with accelerated lambda iteration was employed to carry out the calculations.

\begin{figure}
\resizebox{\hsize}{!}{\includegraphics{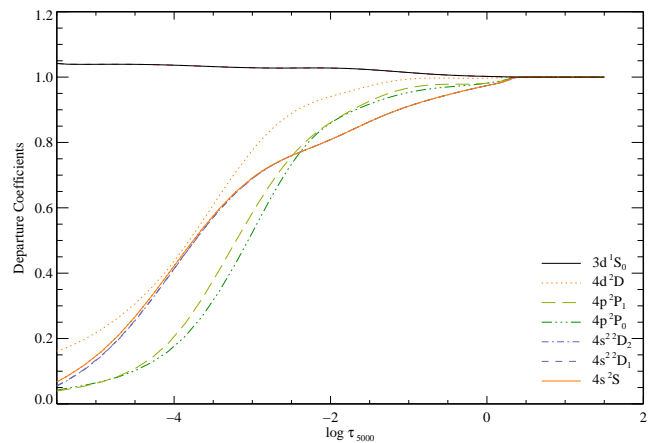}}
\caption{Departure coefficients \bi\ of the selected energy levels (listed in the figure) as a function of continuum optical depth at $5000$\,\AA\ for the model atmosphere of \object{G05-40}. The collision with neutral hydrogen was scaled by a factor of 0.1 following \citet{Shi14}}
\label{fig1}
\end{figure}

Figure \ref{fig1} shows the calculated departure coefficients (\bi) of the selected levels as a function of the continuum optical depth at $5000$\,\AA\ (\optd) for the model atmosphere of \object{G05-40}, which has a temperature and metallicity representative of the sample stars. Here $b_i$ represents the ratio between the number densities given by non-LTE calculations (\nin) and LTE (\nil) calculations. The departure coefficients for the most important levels of \cui\ and the ground state of \cuii\ are shown in the figure. The overionization becomes significant outside the layers with \optd $\simeq 0.5$, depopulating the lower energy levels such as \cu{4}{s}{2}{2}{D}{ }{ } and \cu{4}{p}{ }{2}{P}{o}{ }, and thus leading to an underestimation of the copper abundance from the LTE analysis.

\begin{figure*}
\centering
\includegraphics[width=15cm]{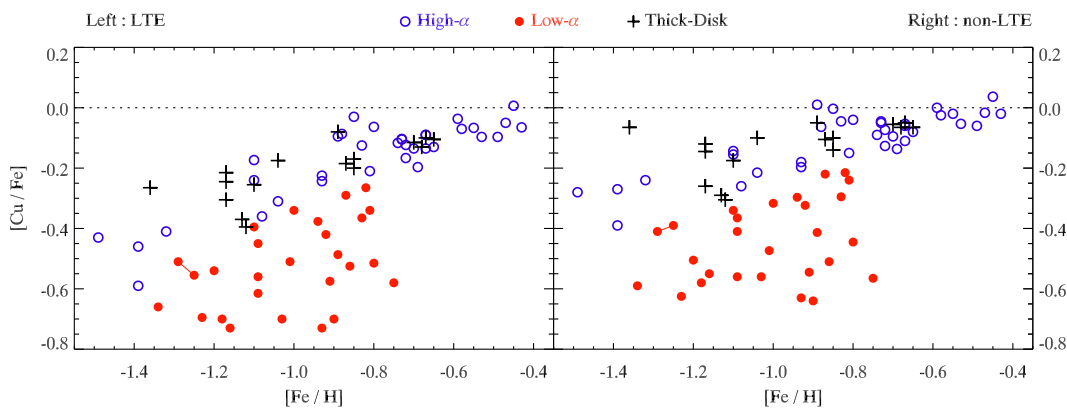}
\caption{Abundance ratios [Cu/Fe] as a function of [Fe/H] for the program stars, where LTE and non-LTE results are represented in the left and right panel, respectively. The high-\alp \ (open
blue circles), low-\alp \ (filled red circles) and thick-disk populations (black crosses) are indicated (see the online version for color copies of the figures). \object{G112-42} and \object{G112-43} are connected with a straight line.}
\label{fig2}
\end{figure*}

\subsection{Copper abundance analysis}
\label{sec2.2}
We investigated 94 dwarfs with effective temperatures between $5200$ K and $6300$ K and a metallicity range from $-1.6$ to $-0.4$, of which 78 belong to the halo population and the remaining 16 are from the thick disk. The stars were observed using either the ESO/VLT UVES spectrograph (with a resolution $R \simeq 55\,000$ and a signal-to-noise ratio $S/N \sim 250-500$) or the NOT/FIES spectrograph ($R \simeq 40\,000$, $S/N \simeq 140-200$). Details about the sample and observations are reported in NS10 and \cite{Sch06}. In addition, we also directly employed the stellar parameters from NS10.

In our copper abundance analysis, we applied the MAFAGS opacity sampling (OS) model atmosphere developed and updated by \citet{Gru04} and \citet{Gru09}. This is a one-dimensional plane-parallel model that is in hydrostatic equilibrium with chemical homogeneity throughout all the atmospheric layers. The turbulent convection is treated with the improved model presented by \citet{Can92}. $86\,000$ frequency points are randomly sampled, including both continuous opacities and line opacities which are calculated based on the line list of \cite{Kur09}. The model was also employed in several previous non-LTE works \citep{Mas11,Shi14, Yan15}. The differences between MAFAGS and MARCS OS model have been discussed in \cite{Shi14}, and no significant departure was found. 

For the fully differential abundance evaluation with respect to the Sun, three lines are employed to perform the spectrum synthesis, which are $5105.5$\,\AA, $5218.2$\,\AA, and $5782.1$\,\AA. The $\lambda$ $5782.1$ is not available in the UVES spectra as mentioned in NS11, so it only takes a part in the analysis of the FIES spectra. The oscillator strengths ($\log gf$) used in this work were calibrated with respect to the Sun as described by \citet{Shi14}. The transitions of hyperfine structure (HFS) were computed based on the Russell-Saunders (RS) coupling method, with the input parameters being adopted according to \citet{Bie76}. The van der Waals damping constants ($\log C_6$) for \cui\ lines were obtained based on \citet{Ans91, Ans95}. The transitions and the relevant line data are presented in Table 1 of \citet{Yan15}.

We used the Spectrum Investigation Utility (SIU) developed by \citet{Ree91} to perform the line formation. In the program, the external broadening mechanisms are combined together as one Gauss profile to be convolved with the synthetic spectrum during the line profile fitting. In addition, the isotope ratio of copper ($^{63}$Cu and $^{65}$Cu) was set to be $0.69:0.31$ following \citet{Asp09} (see Fig.\ref{fig4} for an example of the fitting of the $5105.5$\,\AA\ line). We adopted $\eps {(Cu)} = 4.25$ \citep{Lod09} as the ``absolute'' solar copper abundance, which is measured from meteorites.

\section{Results}
\label{sec3}
Similar to NS11, the copper abundances were derived for 78 stars of the sample. The remaining 16 stars do not have recognizable or clear enough \cui\ line profiles in their spectra. We present our detailed results in Tables \ref{tab1} and \ref{tab2}, which include both LTE and non-LTE copper abundances derived from individual lines for each star. The stellar \cufe\ was assigned by computing the arithmetical mean value of all the valid line results for each star, and the error given in the tables is derived from line-to-line scatter. As seen from Table \ref{tab1} and Table \ref{tab2}, there is no large line-to-line scatter for the copper abundance (see also Fig.\ref{fig5}).  Figure \ref{fig2} shows our final stellar \cufe\ obtained from LTE (left panel) and non-LTE calculations (right panel) as a function of \feh. For a better comparison, the coordinate scales in Figure \ref{fig2} are exactly the same in both panels. The high-\alp, low-\alp\text{,} and thick-disk populations are indicated by different symbols. \object{G112-42} and \object{G112-43} are probably components of a binary system, therefore we connect them with a straight line in the figure the same as NS11. 

Our LTE \cufe\ trend (Fig. \ref{fig2} left panel) agrees quite well with that in NS11. The stars that have higher \alpfe\ tend to have higher \cufe\ compared to the low-\alp \ stars in the same metallicity range. The high-\alp \ stars also show a tightly increasing \cufe\ trend toward higher metallicity, which overlaps with the thick-disk stars. In contrast, the low-\alp \ stars lie below the other two populations in the figure, and they present a much looser relation between \cufe\ and \feh, showing large scatters at a given metallicity. The consistency of the two works is expected because we adopted the same stellar parameters and similar $\log gf$ values. Still, the agreement is remarkable (see Fig. \ref{fig6}) considering that NS11 applied equivalent widths to derive Cu abundances, whereas we used profile fitting. 

When considering the non-LTE results (Fig. \ref{fig2} right panel), it is still tenable that  high- and low-\alp \ stars can be separated by their \cufe\ values, suggesting they may have different origins and evolution histories, as discussed in NS10 and NS11. This separation is even clearer in the lower metallicity range for non-LTE results, namely from  $-1.6$ to $-1.1$. The clearer separation is mainly because the non-LTE effects for copper are more sensitive to low metallicity, which significantly reduces the number density of the free electrons in the stellar atmosphere, leading to a lower collisional rate that is not efficient enough for the environment to reach the LTE state. 

\begin{figure}
\centering
\resizebox{8cm}{!}
{\includegraphics{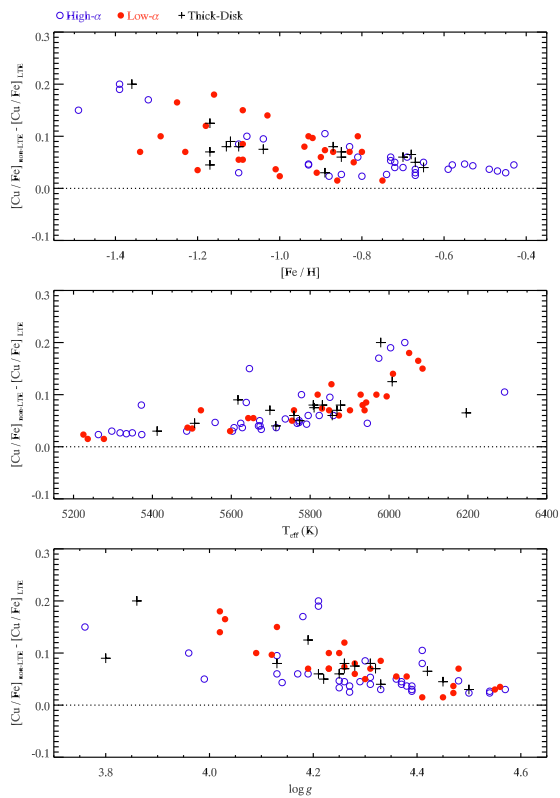}}
\caption{Differences in \cufe between non-LTE and LTE as a function of metallicity (top panel), effective temperature (middle panel), and surface gravity (bottom panel). The symbols are the same as in Fig. \ref{fig2}}
\label{fig3}
\end{figure}

The non-LTE corrections are dominated by the properties of the energy levels and the transitions between them, and consequently, they differ from line to line. In general, the non-LTE effects tend to increase the copper abundances when \feh $< 0$ (Tables \ref{tab1} and \ref{tab2}). \citet{Yan15} found that the LTE analyses underestimated the copper abundances by $0.01$ to $0.2$ dex depending on the stellar parameters; similar results can be seen in this work. In addition, the departure from LTE changes the \cufe\ trend significantly at the low-metallicity end. A much flatter \cufe\ distribution in the metallicity range $-1.6<\feh<-0.8$ is shown in Fig. \ref{fig2} (compared with the left panel) for thick-disk and halo stars, a linear least-squares fitting gives a slope of 0.54 to LTE results and 0.33 to non-LTE results. The non-LTE departures as functions of stellar \feh, \teff, and \logg\ are shown in Fig. \ref{fig3} (top, middle, and bottom panel, respectively). The top panel shows an increasing trend toward the lower metallicity, which is also presented by \citet{Yan15}. This dependence affects the \cufe\ values in metal-poor stars much more than that in metal-rich stars and thus flattens the \cufe\ versus \feh\ relation. The non-LTE effects tend to be strong for hot stars with two outsiders whose temperatures are the highest in our sample (middle panel). At higher temperature, \cui\ starts to become a minority species, thus small corrections from statistical equilibrium to the neutral copper can lead to large abundance deviations, but this is not the case if the temperature continues to increase because the \cui\ lines become weak; they form in a much deeper layer of a stellar atmosphere, where non-LTE effects are weaker (see Fig. \ref{fig1}). The non-LTE departure seems loosely dependent on the surface gravities (bottom panel), while this relation is not seen in our previous work \citep{Yan15}. The dependence is probably spurious because the three stars with the lowest surface gravities are all metal-poor stars (see Fig. \ref{fig7}).

\section{Conclusions}
\label{sec4}
We analyzed the copper abundances for the 94 stars from \citet{Nis10} with \teff\ between $5200$ K and $6300$ K and \feh\ from $-1.6$ to $-0.4$ using both LTE and non-LTE calculations. Based on the results, we summarize our conclusions as follows.

\begin{enumerate}
\item Two distinct halo populations separated by \alpfe\ still show distinguishable \cufe\ in our non-LTE results. The separation of \cufe\ between high- and low-\alp \ stars is even clearer, especially for the metal-poor stars. The deficient copper abundances in the low-\alp \ sample suggests that they may originate from dwarf galaxies that were accreted into the Milky Way. 
\item The non-LTE effects are strong for copper, especially in the metal-poor stars. The copper abundances can be underestimated by up to 0.2 dex by LTE analysis. 
\item The non-LTE corrections differ from line to line. The departure from LTE shows clear dependence on the stellar metallicity, leading to a much flatter distribution with \feh\ in the metallicity range $-1.6$ to $-0.8$ for the high-\alp \ and thick-disk stars, with a slope of 0.54 for LTE results and 0.33 for non-LTE results. 
\end{enumerate}

\begin{acknowledgements}
This research was supported by National Key Basic Research Program of China 2014CB845700, and by the National Natural Science Foundation of China under grant Nos. 11321064, 11233004, 11390371, 11473033, U1331122. 
P. E. N. acknowledges a visiting professorship at the National Astronomical Observatories in Beijing granted by the Chinese Academy of Sciences (Contract no. 6-1309001).  Funding for the Stellar Astrophysics Centre is provided by the Danish National Research Foundation (Grant agreement no.: DNRF106). 
\end{acknowledgements}

\Online

\begin{figure}
\resizebox{\hsize}{!}{\includegraphics{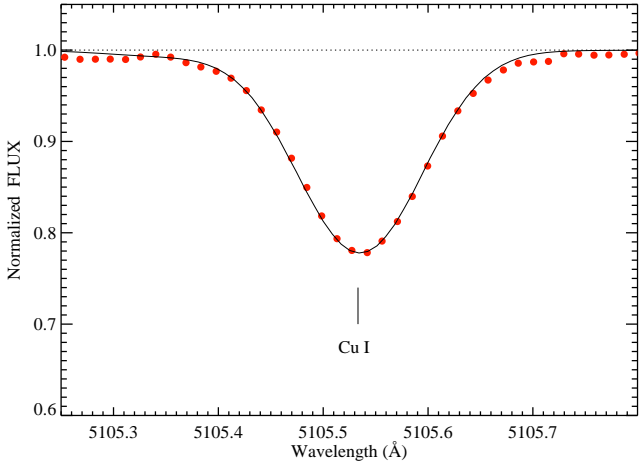}}
\caption{Synthetic profile of the \cui\ $5105$\,\AA\ line for \object{G05-40} from UVES spectra. The observed spectrum and theoretical synthesis are represented by red filled circles and the solid line, respectively. }
\label{fig4}
\end{figure}

\begin{figure}
\resizebox{\hsize}{!}{\includegraphics{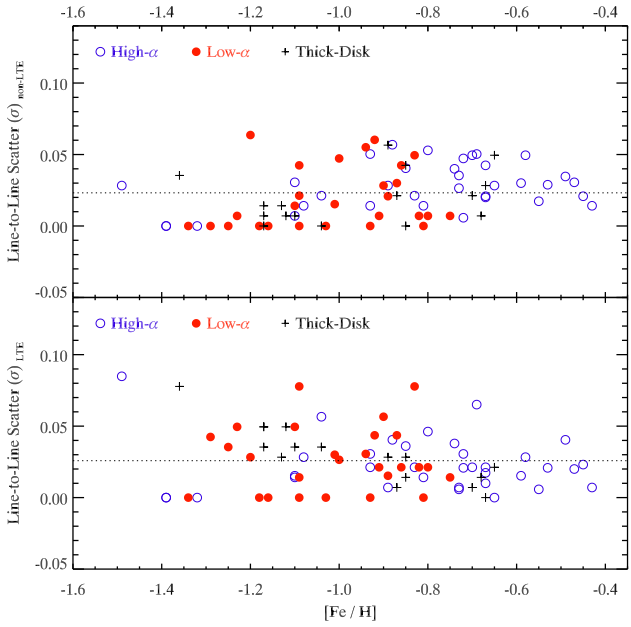}}
\caption{Distribution of line-to-line scatter as a function of metallicity for LTE abundances (bottom panel) and non-LTE abundances (top panel). The stars whose copper abundances were derived by only one line are not plotted in the figure. The symbols are as same as in Fig. \ref{fig2}.}
\label{fig5}
\end{figure}

\begin{figure}
\resizebox{\hsize}{!}{\includegraphics{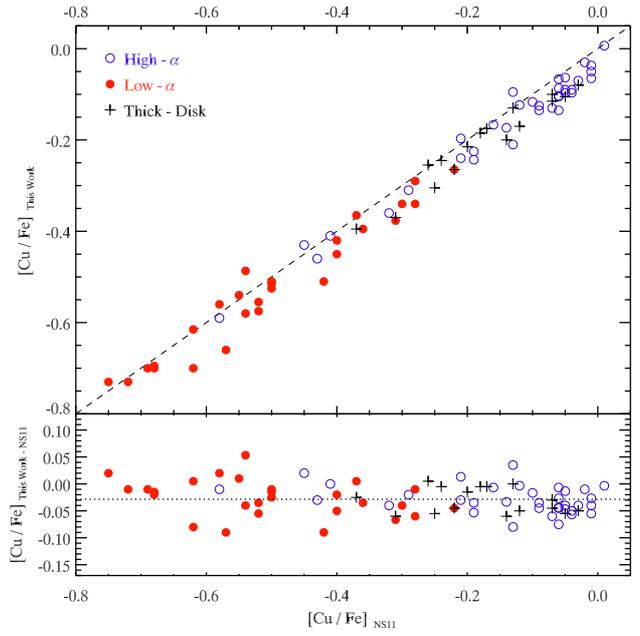}}
\caption{Comparison between this work and NS11 for the LTE results. The symbols are the same as in Fig. \ref{fig2}.}
\label{fig6}
\end{figure}

\begin{figure}
\resizebox{\hsize}{!}{\includegraphics{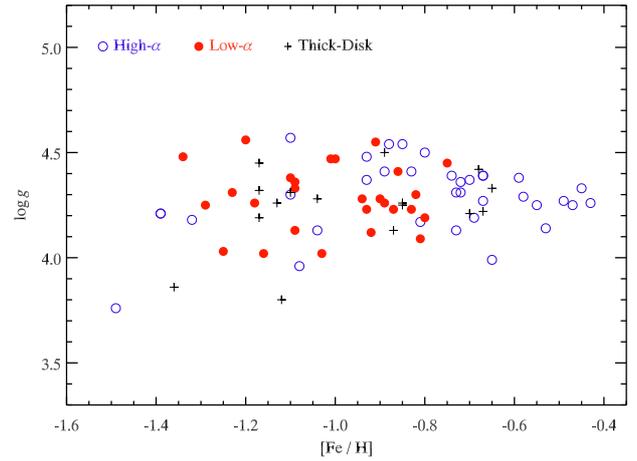}}
\caption{Distribution of surface gravity as a function of metallicity. The symbols are as same as in Fig. \ref{fig2}.}
\label{fig7}
\end{figure}

\begin{table*}
\centering
\caption[1]{LTE and non-LTE copper abundance ratios \cufe\ for stars with VLT/UVES spectra.}
\label{tab1}
\setlength{\tabcolsep}{0.2cm}
\begin{tabular}{lccccrrrrrcc}
\noalign{\smallskip}
\hline\hline
\noalign{\smallskip}
\noalign{\smallskip}
Star & \teff\ (K)  & \logg  & \feh  & $\xi$ & \alpfe & $5105$\AA  & $5218$\AA  & $5782$\AA  & \cufe  & Class & LFS\\
\noalign{\smallskip}
\hline
\noalign{\smallskip}
  BD$-$21 3420  &  $5808$  &  $4.26$  &  $-1.13$  &  $1.30$  &  $0.31$  &  $-0.39$  &  $-0.35$  &  $  ...$  &  $-0.37 \pm 0.03$  &  TD  &  L  \\
                &  $    $  &  $    $  &  $     $  &  $    $  &  $    $  &  $-0.28$  &  $-0.30$  &  $  ...$  &  $-0.29 \pm 0.01$  &      &  N  \\
  CD$-$33 3337  &  $5979$  &  $3.86$  &  $-1.36$  &  $1.70$  &  $0.30$  &  $-0.32$  &  $-0.21$  &  $  ...$  &  $-0.26 \pm 0.08$  &  TD  &  L  \\
                &  $    $  &  $    $  &  $     $  &  $    $  &  $    $  &  $-0.09$  &  $-0.04$  &  $  ...$  &  $-0.06 \pm 0.04$  &      &  N  \\
  CD$-$43 6810  &  $5945$  &  $4.26$  &  $-0.43$  &  $1.30$  &  $0.23$  &  $-0.07$  &  $-0.06$  &  $  ...$  &  $-0.07 \pm 0.01$  &high-\alp&  L  \\
                &  $    $  &  $    $  &  $     $  &  $    $  &  $    $  &  $-0.01$  &  $-0.03$  &  $  ...$  &  $-0.02 \pm 0.01$  &      &  N  \\
  CD$-$45 3283  &  $5597$  &  $4.55$  &  $-0.91$  &  $1.00$  &  $0.12$  &  $-0.59$  &  $-0.56$  &  $  ...$  &  $-0.57 \pm 0.02$  & low-\alp&  L  \\
                &  $    $  &  $    $  &  $     $  &  $    $  &  $    $  &  $-0.54$  &  $-0.55$  &  $  ...$  &  $-0.54 \pm 0.01$  &      &  N  \\
  CD$-$57 1633  &  $5873$  &  $4.28$  &  $-0.90$  &  $1.20$  &  $0.07$  &  $-0.74$  &  $-0.66$  &  $  ...$  &  $-0.70 \pm 0.06$  & low-\alp&  L  \\
                &  $    $  &  $    $  &  $     $  &  $    $  &  $    $  &  $-0.66$  &  $-0.62$  &  $  ...$  &  $-0.64 \pm 0.03$  &      &  N  \\
   CD$-$61 282  &  $5759$  &  $4.31$  &  $-1.23$  &  $1.30$  &  $0.22$  &  $-0.73$  &  $-0.66$  &  $  ...$  &  $-0.69 \pm 0.05$  & low-\alp&  L  \\
                &  $    $  &  $    $  &  $     $  &  $    $  &  $    $  &  $-0.63$  &  $-0.62$  &  $  ...$  &  $-0.63 \pm 0.01$  &      &  N  \\
        G05-19  &  $5854$  &  $4.26$  &  $-1.18$  &  $1.30$  &  $0.19$  &  $-0.70$  &  $  ...$  &  $  ...$  &  $-0.70 \pm 0.00$  & low-\alp&  L  \\
                &  $    $  &  $    $  &  $     $  &  $    $  &  $    $  &  $-0.58$  &  $  ...$  &  $  ...$  &  $-0.58 \pm 0.00$  &      &  N  \\
        G05-40  &  $5795$  &  $4.17$  &  $-0.81$  &  $1.20$  &  $0.31$  &  $-0.22$  &  $-0.20$  &  $  ...$  &  $-0.21 \pm 0.01$  &high-\alp&  L  \\
                &  $    $  &  $    $  &  $     $  &  $    $  &  $    $  &  $-0.14$  &  $-0.16$  &  $  ...$  &  $-0.15 \pm 0.01$  &      &  N  \\
        G18-28  &  $5372$  &  $4.41$  &  $-0.83$  &  $1.00$  &  $0.31$  &  $-0.11$  &  $-0.14$  &  $  ...$  &  $-0.13 \pm 0.02$  &high-\alp&  L  \\
                &  $    $  &  $    $  &  $     $  &  $    $  &  $    $  &  $-0.06$  &  $-0.03$  &  $  ...$  &  $-0.05 \pm 0.02$  &      &  N  \\
        G18-39  &  $6040$  &  $4.21$  &  $-1.39$  &  $1.50$  &  $0.34$  &  $-0.59$  &  $  ...$  &  $  ...$  &  $-0.59 \pm 0.00$  &high-\alp&  L  \\
                &  $    $  &  $    $  &  $     $  &  $    $  &  $    $  &  $-0.39$  &  $  ...$  &  $  ...$  &  $-0.39 \pm 0.00$  &      &  N  \\
        G46-31  &  $5901$  &  $4.23$  &  $-0.83$  &  $1.40$  &  $0.15$  &  $-0.42$  &  $-0.31$  &  $  ...$  &  $-0.37 \pm 0.08$  & low-\alp&  L  \\
                &  $    $  &  $    $  &  $     $  &  $    $  &  $    $  &  $-0.33$  &  $-0.26$  &  $  ...$  &  $-0.30 \pm 0.05$  &      &  N  \\
        G66-22  &  $5236$  &  $4.41$  &  $-0.86$  &  $0.90$  &  $0.12$  &  $-0.51$  &  $-0.54$  &  $  ...$  &  $-0.52 \pm 0.02$  & low-\alp&  L  \\
                &  $    $  &  $    $  &  $     $  &  $    $  &  $    $  &  $-0.48$  &  $-0.54$  &  $  ...$  &  $-0.51 \pm 0.04$  &      &  N  \\
        G82-05  &  $5277$  &  $4.45$  &  $-0.75$  &  $0.90$  &  $0.09$  &  $-0.59$  &  $-0.57$  &  $  ...$  &  $-0.58 \pm 0.01$  & low-\alp&  L  \\
                &  $    $  &  $    $  &  $     $  &  $    $  &  $    $  &  $-0.56$  &  $-0.57$  &  $  ...$  &  $-0.56 \pm 0.01$  &      &  N  \\
       G112-43  &  $6074$  &  $4.03$  &  $-1.25$  &  $1.30$  &  $0.24$  &  $-0.58$  &  $-0.53$  &  $  ...$  &  $-0.55 \pm 0.04$  & low-\alp&  L  \\
                &  $    $  &  $    $  &  $     $  &  $    $  &  $    $  &  $-0.39$  &  $-0.39$  &  $  ...$  &  $-0.39 \pm 0.00$  &      &  N  \\
       G112-44  &  $5819$  &  $4.25$  &  $-1.29$  &  $1.20$  &  $0.22$  &  $-0.54$  &  $-0.48$  &  $  ...$  &  $-0.51 \pm 0.04$  & low-\alp&  L  \\
                &  $    $  &  $    $  &  $     $  &  $    $  &  $    $  &  $-0.41$  &  $-0.41$  &  $  ...$  &  $-0.41 \pm 0.00$  &      &  N  \\
       G114-42  &  $5643$  &  $4.38$  &  $-1.10$  &  $1.30$  &  $0.19$  &  $-0.43$  &  $-0.36$  &  $  ...$  &  $-0.40 \pm 0.05$  & low-\alp&  L  \\
                &  $    $  &  $    $  &  $     $  &  $    $  &  $    $  &  $-0.35$  &  $-0.33$  &  $  ...$  &  $-0.34 \pm 0.01$  &      &  N  \\
       G121-12  &  $5928$  &  $4.23$  &  $-0.93$  &  $1.40$  &  $0.10$  &  $-0.73$  &  $  ...$  &  $  ...$  &  $-0.73 \pm 0.00$  & low-\alp&  L  \\
                &  $    $  &  $    $  &  $     $  &  $    $  &  $    $  &  $-0.63$  &  $  ...$  &  $  ...$  &  $-0.63 \pm 0.00$  &      &  N  \\
       G159-50  &  $5624$  &  $4.37$  &  $-0.93$  &  $1.10$  &  $0.31$  &  $-0.24$  &  $-0.21$  &  $  ...$  &  $-0.22 \pm 0.02$  &high-\alp&  L  \\
                &  $    $  &  $    $  &  $     $  &  $    $  &  $    $  &  $-0.17$  &  $-0.19$  &  $  ...$  &  $-0.18 \pm 0.01$  &      &  N  \\
       G188-22  &  $5974$  &  $4.18$  &  $-1.32$  &  $1.50$  &  $0.35$  &  $-0.41$  &  $  ...$  &  $  ...$  &  $-0.41 \pm 0.00$  &high-\alp&  L  \\
                &  $    $  &  $    $  &  $     $  &  $    $  &  $    $  &  $-0.24$  &  $  ...$  &  $  ...$  &  $-0.24 \pm 0.00$  &      &  N  \\
     HD\,3567  &  $6051$  &  $4.02$  &  $-1.16$  &  $1.50$  &  $0.21$  &  $-0.73$  &  $  ...$  &  $  ...$  &  $-0.73 \pm 0.00$  & low-\alp&  L  \\
                &  $    $  &  $    $  &  $     $  &  $    $  &  $    $  &  $-0.55$  &  $  ...$  &  $  ...$  &  $-0.55 \pm 0.00$  &      &  N  \\
   HD\,17820  &  $5773$  &  $4.22$  &  $-0.67$  &  $1.30$  &  $0.29$  &  $-0.10$  &  $-0.10$  &  $  ...$  &  $-0.10 \pm 0.00$  &  TD  &  L  \\
                &  $    $  &  $    $  &  $     $  &  $    $  &  $    $  &  $-0.03$  &  $-0.07$  &  $  ...$  &  $-0.05 \pm 0.03$  &      &  N  \\
   HD\,22879  &  $5759$  &  $4.25$  &  $-0.85$  &  $1.30$  &  $0.31$  &  $-0.22$  &  $-0.18$  &  $  ...$  &  $-0.20 \pm 0.03$  &  TD  &  L  \\
                &  $    $  &  $    $  &  $     $  &  $    $  &  $    $  &  $-0.14$  &  $-0.14$  &  $  ...$  &  $-0.14 \pm 0.00$  &      &  N  \\
   HD\,25704  &  $5868$  &  $4.26$  &  $-0.85$  &  $1.40$  &  $0.24$  &  $-0.16$  &  $-0.18$  &  $  ...$  &  $-0.17 \pm 0.01$  &  TD  &  L  \\
                &  $    $  &  $    $  &  $     $  &  $    $  &  $    $  &  $-0.07$  &  $-0.13$  &  $  ...$  &  $-0.10 \pm 0.04$  &      &  N  \\
   HD\,51754  &  $5767$  &  $4.29$  &  $-0.58$  &  $1.40$  &  $0.26$  &  $-0.05$  &  $-0.09$  &  $  ...$  &  $-0.07 \pm 0.03$  &high-\alp&  L  \\
                &  $    $  &  $    $  &  $     $  &  $    $  &  $    $  &  $ 0.01$  &  $-0.06$  &  $  ...$  &  $-0.03 \pm 0.05$  &      &  N  \\
   HD\,76932  &  $5877$  &  $4.13$  &  $-0.87$  &  $1.40$  &  $0.29$  &  $-0.19$  &  $-0.18$  &  $  ...$  &  $-0.18 \pm 0.01$  &  TD  &  L  \\
                &  $    $  &  $    $  &  $     $  &  $    $  &  $    $  &  $-0.09$  &  $-0.12$  &  $  ...$  &  $-0.11 \pm 0.02$  &      &  N  \\
   HD\,97320  &  $6008$  &  $4.19$  &  $-1.17$  &  $1.60$  &  $0.28$  &  $-0.28$  &  $-0.21$  &  $  ...$  &  $-0.24 \pm 0.05$  &  TD  &  L  \\
                &  $    $  &  $    $  &  $     $  &  $    $  &  $    $  &  $-0.13$  &  $-0.11$  &  $  ...$  &  $-0.12 \pm 0.01$  &      &  N  \\
 HD\,103723  &  $5938$  &  $4.19$  &  $-0.80$  &  $1.20$  &  $0.14$  &  $-0.53$  &  $-0.50$  &  $  ...$  &  $-0.51 \pm 0.02$  & low-\alp&  L  \\
                &  $    $  &  $    $  &  $     $  &  $    $  &  $    $  &  $-0.44$  &  $-0.45$  &  $  ...$  &  $-0.45 \pm 0.01$  &      &  N  \\
 HD\,105004  &  $5754$  &  $4.30$  &  $-0.82$  &  $1.20$  &  $0.14$  &  $-0.28$  &  $-0.25$  &  $  ...$  &  $-0.27 \pm 0.02$  & low-\alp&  L  \\
                &  $    $  &  $    $  &  $     $  &  $    $  &  $    $  &  $-0.21$  &  $-0.22$  &  $  ...$  &  $-0.22 \pm 0.01$  &      &  N  \\
   HD\,106516  &  $6196$  &  $4.42$  &  $-0.68$  &  $1.30$  &  $0.29$  &  $-0.14$  &  $-0.12$  &  $  ...$  &  $-0.13 \pm 0.01$  &  TD  &  L  \\
                &  $    $  &  $    $  &  $     $  &  $    $  &  $    $  &  $-0.06$  &  $-0.07$  &  $  ...$  &  $-0.06 \pm 0.01$  &      &  N  \\
\noalign{\smallskip}

\end{tabular}
\end{table*}

\begin{table*}
\centering
Table \ref{tab1} continued.
\setlength{\tabcolsep}{0.2cm}
\begin{tabular}{lccccrrrrrcc}
\noalign{\smallskip}
\hline\hline
\noalign{\smallskip}
\noalign{\smallskip}
Star & \teff\ (K)  & \logg  & \feh  & $\xi$ & \alpfe & $5105$\AA  & $5218$\AA  & $5782$\AA  & \cufe  & Class & LFS\\
\noalign{\smallskip}
\hline
\noalign{\smallskip}
    HD\,111980  &  $5778$  &  $3.96$  &  $-1.08$  &  $1.50$  &  $0.34$  &  $-0.38$  &  $-0.34$  &  $  ...$  &  $-0.36 \pm 0.03$  &high-\alp&  L  \\
                &  $    $  &  $    $  &  $     $  &  $    $  &  $    $  &  $-0.25$  &  $-0.27$  &  $  ...$  &  $-0.26 \pm 0.01$  &      &  N  \\
    HD\,113679  &  $5672$  &  $3.99$  &  $-0.65$  &  $1.40$  &  $0.32$  &  $-0.13$  &  $-0.13$  &  $  ...$  &  $-0.13 \pm 0.00$  &high-\alp&  L  \\
                &  $    $  &  $    $  &  $     $  &  $    $  &  $    $  &  $-0.06$  &  $-0.10$  &  $  ...$  &  $-0.08 \pm 0.03$  &      &  N  \\
  HD\,114762A  &  $5856$  &  $4.21$  &  $-0.70$  &  $1.50$  &  $0.24$  &  $-0.12$  &  $-0.11$  &  $  ...$  &  $-0.11 \pm 0.01$  &  TD  &  L  \\
                &  $    $  &  $    $  &  $     $  &  $    $  &  $    $  &  $-0.04$  &  $-0.07$  &  $  ...$  &  $-0.06 \pm 0.02$  &      &  N  \\
    HD\,120559  &  $5412$  &  $4.50$  &  $-0.89$  &  $1.10$  &  $0.30$  &  $-0.06$  &  $-0.10$  &  $  ...$  &  $-0.08 \pm 0.03$  &  TD  &  L  \\
                &  $    $  &  $    $  &  $     $  &  $    $  &  $    $  &  $-0.01$  &  $-0.09$  &  $  ...$  &  $-0.05 \pm 0.06$  &      &  N  \\
    HD\,121004  &  $5669$  &  $4.37$  &  $-0.70$  &  $1.30$  &  $0.32$  &  $-0.12$  &  $-0.15$  &  $  ...$  &  $-0.13 \pm 0.02$  &high-\alp&  L  \\
                &  $    $  &  $    $  &  $     $  &  $    $  &  $    $  &  $-0.06$  &  $-0.13$  &  $  ...$  &  $-0.09 \pm 0.05$  &      &  N  \\
    HD\,126681  &  $5507$  &  $4.45$  &  $-1.17$  &  $1.20$  &  $0.35$  &  $-0.33$  &  $-0.28$  &  $  ...$  &  $-0.31 \pm 0.04$  &  TD  &  L  \\
                &  $    $  &  $    $  &  $     $  &  $    $  &  $    $  &  $-0.26$  &  $-0.26$  &  $  ...$  &  $-0.26 \pm 0.00$  &      &  N  \\
    HD\,132475  &  $5646$  &  $3.76$  &  $-1.49$  &  $1.50$  &  $0.38$  &  $-0.49$  &  $-0.37$  &  $  ...$  &  $-0.43 \pm 0.08$  &? high-\alp&  L  \\
                &  $    $  &  $    $  &  $     $  &  $    $  &  $    $  &  $-0.30$  &  $-0.26$  &  $  ...$  &  $-0.28 \pm 0.03$  &      &  N  \\
    HD\,148816  &  $5823$  &  $4.13$  &  $-0.73$  &  $1.40$  &  $0.27$  &  $-0.10$  &  $-0.11$  &  $  ...$  &  $-0.10 \pm 0.01$  &high-\alp&  L  \\
                &  $    $  &  $    $  &  $     $  &  $    $  &  $    $  &  $-0.02$  &  $-0.07$  &  $  ...$  &  $-0.05 \pm 0.04$  &      &  N  \\
    HD\,163810  &  $5501$  &  $4.56$  &  $-1.20$  &  $1.30$  &  $0.21$  &  $-0.52$  &  $-0.56$  &  $  ...$  &  $-0.54 \pm 0.03$  & low-\alp&  L  \\
                &  $    $  &  $    $  &  $     $  &  $    $  &  $    $  &  $-0.46$  &  $-0.55$  &  $  ...$  &  $-0.50 \pm 0.06$  &      &  N  \\
    HD\,175179  &  $5713$  &  $4.33$  &  $-0.65$  &  $1.20$  &  $0.29$  &  $-0.09$  &  $-0.12$  &  $  ...$  &  $-0.11 \pm 0.02$  &  TD  &  L  \\
                &  $    $  &  $    $  &  $     $  &  $    $  &  $    $  &  $-0.03$  &  $-0.10$  &  $  ...$  &  $-0.06 \pm 0.05$  &      &  N  \\
    HD\,179626  &  $5850$  &  $4.13$  &  $-1.04$  &  $1.60$  &  $0.31$  &  $-0.35$  &  $-0.27$  &  $  ...$  &  $-0.31 \pm 0.06$  &high-\alp&  L  \\
                &  $    $  &  $    $  &  $     $  &  $    $  &  $    $  &  $-0.23$  &  $-0.20$  &  $  ...$  &  $-0.21 \pm 0.02$  &      &  N  \\
    HD\,189558  &  $5617$  &  $3.80$  &  $-1.12$  &  $1.40$  &  $0.33$  &  $-0.43$  &  $-0.36$  &  $  ...$  &  $-0.40 \pm 0.05$  &  TD  &  L  \\
                &  $    $  &  $    $  &  $     $  &  $    $  &  $    $  &  $-0.31$  &  $-0.30$  &  $  ...$  &  $-0.31 \pm 0.01$  &      &  N  \\
    HD\,193901  &  $5656$  &  $4.36$  &  $-1.09$  &  $1.20$  &  $0.16$  &  $-0.67$  &  $-0.56$  &  $  ...$  &  $-0.62 \pm 0.08$  & low-\alp&  L  \\
                &  $    $  &  $    $  &  $     $  &  $    $  &  $    $  &  $-0.59$  &  $-0.53$  &  $  ...$  &  $-0.56 \pm 0.04$  &      &  N  \\
    HD\,194598  &  $5942$  &  $4.33$  &  $-1.09$  &  $1.40$  &  $0.18$  &  $-0.46$  &  $-0.44$  &  $  ...$  &  $-0.45 \pm 0.01$  & low-\alp&  L  \\
                &  $    $  &  $    $  &  $     $  &  $    $  &  $    $  &  $-0.35$  &  $-0.38$  &  $  ...$  &  $-0.37 \pm 0.02$  &      &  N  \\
    HD\,199289  &  $5810$  &  $4.28$  &  $-1.04$  &  $1.30$  &  $0.30$  &  $-0.20$  &  $-0.15$  &  $  ...$  &  $-0.17 \pm 0.04$  &  TD  &  L  \\
                &  $    $  &  $    $  &  $     $  &  $    $  &  $    $  &  $-0.10$  &  $-0.10$  &  $  ...$  &  $-0.10 \pm 0.00$  &      &  N  \\
    HD\,205650  &  $5698$  &  $4.32$  &  $-1.17$  &  $1.30$  &  $0.30$  &  $-0.25$  &  $-0.18$  &  $  ...$  &  $-0.21 \pm 0.05$  &  TD  &  L  \\
                &  $    $  &  $    $  &  $     $  &  $    $  &  $    $  &  $-0.15$  &  $-0.14$  &  $  ...$  &  $-0.14 \pm 0.01$  &      &  N  \\
    HD\,222766  &  $5334$  &  $4.27$  &  $-0.67$  &  $0.80$  &  $0.30$  &  $-0.12$  &  $-0.15$  &  $  ...$  &  $-0.13 \pm 0.02$  &high-\alp&  L  \\
                &  $    $  &  $    $  &  $     $  &  $    $  &  $    $  &  $-0.08$  &  $-0.14$  &  $  ...$  &  $-0.11 \pm 0.04$  &      &  N  \\
    HD\,241253  &  $5831$  &  $4.31$  &  $-1.10$  &  $1.30$  &  $0.29$  &  $-0.28$  &  $-0.23$  &  $  ...$  &  $-0.25 \pm 0.04$  &  TD  &  L  \\
                &  $    $  &  $    $  &  $     $  &  $    $  &  $    $  &  $-0.17$  &  $-0.18$  &  $  ...$  &  $-0.17 \pm 0.01$  &      &  N  \\
\noalign{\smallskip}
\hline
\end{tabular}
\tablefoot{The stellar parameters including \alpfe\ are the same as \citet{Nis10}. Cols. $7$, $8$, and $9$ are the abundance ratios derived from corresponding \cui\ lines, and Col. $10$ presents both stellar LTE and non-LTE copper abundance ratios (for each star, first and second row, respectively), and the line-to-line scatter is also provided in this column (following the `$\pm$' sign). The classification is indicated by `TD' (thick-disk), `high-\alp' and `low-\alp', and a `?' are added if the classification is uncertain. The rightmost column shows the line formation scenario for each star, where 'L' represents LTE line formation and 'N' represents non-LTE line formation. The stars without available copper abundance are not listed in the table.}
\end{table*}

\begin{table*}
\centering
\caption[2]{LTE and non-LTE copper abundance ratios \cufe\ for stars with NOT/FIES spectra.}
\label{tab2}
\setlength{\tabcolsep}{0.2cm}
\begin{tabular}{lccccrrrrrcc}
\noalign{\smallskip}
\hline\hline
\noalign{\smallskip}
\noalign{\smallskip}
Star & \teff\ (K)  & \logg  & \feh  & $\xi$ & \alpfe & $5105$\AA  & $5218$\AA  & $5782$\AA  & \cufe  & Class & LFS\\
\noalign{\smallskip}
\hline
\noalign{\smallskip}
        G13-38  &  $5263$  &  $4.54$  &  $-0.88$  &  $0.90$  &  $0.32$  &  $-0.04$  &  $-0.11$  &  $-0.11$  &  $-0.09 \pm 0.04$  &high-\alp&  L  \\
                &  $    $  &  $    $  &  $     $  &  $    $  &  $    $  &  $ 0.00$  &  $-0.11$  &  $-0.08$  &  $-0.06 \pm 0.06$  &      &  N  \\
        G15-23  &  $5297$  &  $4.57$  &  $-1.10$  &  $1.00$  &  $0.34$  &  $-0.16$  &  $-0.17$  &  $-0.19$  &  $-0.17 \pm 0.02$  &high-\alp&  L  \\
                &  $    $  &  $    $  &  $     $  &  $    $  &  $    $  &  $-0.11$  &  $-0.17$  &  $-0.15$  &  $-0.14 \pm 0.03$  &      &  N  \\
        G24-13  &  $5673$  &  $4.31$  &  $-0.72$  &  $1.00$  &  $0.29$  &  $-0.16$  &  $-0.20$  &  $-0.14$  &  $-0.17 \pm 0.03$  &high-\alp&  L  \\
                &  $    $  &  $    $  &  $     $  &  $    $  &  $    $  &  $-0.11$  &  $-0.18$  &  $-0.09$  &  $-0.13 \pm 0.05$  &      &  N  \\
        G31-55  &  $5638$  &  $4.30$  &  $-1.10$  &  $1.40$  &  $0.29$  &  $-0.25$  &  $  ...$  &  $-0.23$  &  $-0.24 \pm 0.01$  &high-\alp&  L  \\
                &  $    $  &  $    $  &  $     $  &  $    $  &  $    $  &  $-0.16$  &  $  ...$  &  $-0.15$  &  $-0.16 \pm 0.01$  &      &  N  \\
        G49-19  &  $5772$  &  $4.25$  &  $-0.55$  &  $1.20$  &  $0.27$  &  $-0.07$  &  $-0.07$  &  $-0.06$  &  $-0.07 \pm 0.01$  &high-\alp&  L  \\
                &  $    $  &  $    $  &  $     $  &  $    $  &  $    $  &  $-0.01$  &  $-0.04$  &  $-0.01$  &  $-0.02 \pm 0.02$  &      &  N  \\
        G56-30  &  $5830$  &  $4.26$  &  $-0.89$  &  $1.30$  &  $0.11$  &  $-0.50$  &  $-0.47$  &  $-0.49$  &  $-0.49 \pm 0.02$  & low-\alp&  L  \\
                &  $    $  &  $    $  &  $     $  &  $    $  &  $    $  &  $-0.42$  &  $-0.43$  &  $-0.39$  &  $-0.41 \pm 0.02$  &      &  N  \\
        G56-36  &  $5933$  &  $4.28$  &  $-0.94$  &  $1.40$  &  $0.20$  &  $-0.37$  &  $-0.41$  &  $-0.35$  &  $-0.38 \pm 0.03$  & low-\alp&  L  \\
                &  $    $  &  $    $  &  $     $  &  $    $  &  $    $  &  $-0.27$  &  $-0.36$  &  $-0.26$  &  $-0.30 \pm 0.06$  &      &  N  \\
        G57-07  &  $5676$  &  $4.25$  &  $-0.47$  &  $1.10$  &  $0.31$  &  $-0.05$  &  $-0.07$  &  $-0.03$  &  $-0.05 \pm 0.02$  &high-\alp&  L  \\
                &  $    $  &  $    $  &  $     $  &  $    $  &  $    $  &  $-0.01$  &  $-0.05$  &  $ 0.01$  &  $-0.02 \pm 0.03$  &      &  N  \\
        G74-32  &  $5772$  &  $4.36$  &  $-0.72$  &  $1.10$  &  $0.30$  &  $-0.14$  &  $-0.10$  &  $-0.13$  &  $-0.12 \pm 0.02$  &high-\alp&  L  \\
                &  $    $  &  $    $  &  $     $  &  $    $  &  $    $  &  $-0.08$  &  $-0.07$  &  $-0.07$  &  $-0.07 \pm 0.01$  &      &  N  \\
        G75-31  &  $6010$  &  $4.02$  &  $-1.03$  &  $1.40$  &  $0.20$  &  $-0.70$  &  $  ...$  &  $  ...$  &  $-0.70 \pm 0.00$  & low-\alp&  L  \\
                &  $    $  &  $    $  &  $     $  &  $    $  &  $    $  &  $-0.56$  &  $  ...$  &  $  ...$  &  $-0.56 \pm 0.00$  &      &  N  \\
        G81-02  &  $5859$  &  $4.19$  &  $-0.69$  &  $1.30$  &  $0.19$  &  $-0.20$  &  $-0.13$  &  $-0.26$  &  $-0.20 \pm 0.07$  &high-\alp&  L  \\
                &  $    $  &  $    $  &  $     $  &  $    $  &  $    $  &  $-0.13$  &  $-0.09$  &  $-0.19$  &  $-0.14 \pm 0.05$  &      &  N  \\
        G85-13  &  $5628$  &  $4.38$  &  $-0.59$  &  $1.00$  &  $0.28$  &  $-0.02$  &  $-0.05$  &  $-0.04$  &  $-0.04 \pm 0.02$  &high-\alp&  L  \\
                &  $    $  &  $    $  &  $     $  &  $    $  &  $    $  &  $ 0.03$  &  $-0.03$  &  $ 0.00$  &  $ 0.00 \pm 0.03$  &      &  N  \\
        G87-13  &  $6085$  &  $4.13$  &  $-1.09$  &  $1.50$  &  $0.20$  &  $-0.56$  &  $  ...$  &  $  ...$  &  $-0.56 \pm 0.00$  & low-\alp&  L  \\
                &  $    $  &  $    $  &  $     $  &  $    $  &  $    $  &  $-0.41$  &  $  ...$  &  $  ...$  &  $-0.41 \pm 0.00$  &      &  N  \\
        G94-49  &  $5373$  &  $4.50$  &  $-0.80$  &  $1.10$  &  $0.31$  &  $-0.01$  &  $-0.09$  &  $-0.09$  &  $-0.06 \pm 0.05$  &high-\alp&  L  \\
                &  $    $  &  $    $  &  $     $  &  $    $  &  $    $  &  $ 0.02$  &  $-0.08$  &  $-0.06$  &  $-0.04 \pm 0.05$  &      &  N  \\
        G96-20  &  $6293$  &  $4.41$  &  $-0.89$  &  $1.50$  &  $0.28$  &  $-0.09$  &  $-0.10$  &  $  ...$  &  $-0.09 \pm 0.01$  &high-\alp&  L  \\
                &  $    $  &  $    $  &  $     $  &  $    $  &  $    $  &  $ 0.03$  &  $-0.01$  &  $  ...$  &  $ 0.01 \pm 0.03$  &      &  N  \\
        G98-53  &  $5848$  &  $4.23$  &  $-0.87$  &  $1.30$  &  $0.19$  &  $-0.34$  &  $-0.26$  &  $-0.27$  &  $-0.29 \pm 0.04$  & low-\alp&  L  \\
                &  $    $  &  $    $  &  $     $  &  $    $  &  $    $  &  $-0.25$  &  $-0.22$  &  $-0.19$  &  $-0.22 \pm 0.03$  &      &  N  \\
        G99-21  &  $5487$  &  $4.39$  &  $-0.67$  &  $0.90$  &  $0.29$  &  $-0.08$  &  $-0.09$  &  $-0.10$  &  $-0.09 \pm 0.01$  &high-\alp&  L  \\
                &  $    $  &  $    $  &  $     $  &  $    $  &  $    $  &  $-0.04$  &  $-0.08$  &  $-0.06$  &  $-0.06 \pm 0.02$  &      &  N  \\
       G127-26  &  $5791$  &  $4.14$  &  $-0.53$  &  $1.20$  &  $0.24$  &  $-0.08$  &  $-0.09$  &  $-0.12$  &  $-0.10 \pm 0.02$  &high-\alp&  L  \\
                &  $    $  &  $    $  &  $     $  &  $    $  &  $    $  &  $-0.02$  &  $-0.07$  &  $-0.07$  &  $-0.05 \pm 0.03$  &      &  N  \\
       G150-40  &  $5968$  &  $4.09$  &  $-0.81$  &  $1.40$  &  $0.16$  &  $-0.34$  &  $  ...$  &  $  ...$  &  $-0.34 \pm 0.00$  & low-\alp&  L  \\
                &  $    $  &  $    $  &  $     $  &  $    $  &  $    $  &  $-0.24$  &  $  ...$  &  $  ...$  &  $-0.24 \pm 0.00$  &      &  N  \\
       G170-56  &  $5994$  &  $4.12$  &  $-0.92$  &  $1.50$  &  $0.17$  &  $-0.44$  &  $-0.45$  &  $-0.37$  &  $-0.42 \pm 0.04$  & low-\alp&  L  \\
                &  $    $  &  $    $  &  $     $  &  $    $  &  $    $  &  $-0.33$  &  $-0.38$  &  $-0.26$  &  $-0.32 \pm 0.06$  &      &  N  \\
       G172-61  &  $5225$  &  $4.47$  &  $-1.00$  &  $0.90$  &  $0.19$  &  $-0.32$  &  $-0.37$  &  $-0.33$  &  $-0.34 \pm 0.03$  & low-\alp&  L  \\
                &  $    $  &  $    $  &  $     $  &  $    $  &  $    $  &  $-0.28$  &  $-0.37$  &  $-0.30$  &  $-0.32 \pm 0.05$  &      &  N  \\
       G176-53  &  $5523$  &  $4.48$  &  $-1.34$  &  $1.00$  &  $0.18$  &  $-0.66$  &  $  ...$  &  $  ...$  &  $-0.66 \pm 0.00$  & low-\alp&  L  \\
                &  $    $  &  $    $  &  $     $  &  $    $  &  $    $  &  $-0.59$  &  $  ...$  &  $  ...$  &  $-0.59 \pm 0.00$  &      &  N  \\
       G180-24  &  $6004$  &  $4.21$  &  $-1.39$  &  $1.50$  &  $0.33$  &  $-0.46$  &  $  ...$  &  $  ...$  &  $-0.46 \pm 0.00$  &high-\alp&  L  \\
                &  $    $  &  $    $  &  $     $  &  $    $  &  $    $  &  $-0.27$  &  $  ...$  &  $  ...$  &  $-0.27 \pm 0.00$  &      &  N  \\
       G187-18  &  $5607$  &  $4.39$  &  $-0.67$  &  $1.10$  &  $0.26$  &  $-0.08$  &  $-0.08$  &  $-0.11$  &  $-0.09 \pm 0.02$  &high-\alp&  L  \\
                &  $    $  &  $    $  &  $     $  &  $    $  &  $    $  &  $-0.03$  &  $-0.06$  &  $-0.07$  &  $-0.05 \pm 0.02$  &      &  N  \\
       G232-18  &  $5559$  &  $4.48$  &  $-0.93$  &  $1.30$  &  $0.32$  &  $-0.25$  &  $-0.27$  &  $-0.21$  &  $-0.24 \pm 0.03$  &high-\alp&  L  \\
                &  $    $  &  $    $  &  $     $  &  $    $  &  $    $  &  $-0.19$  &  $-0.25$  &  $-0.15$  &  $-0.20 \pm 0.05$  &      &  N  \\
  HD\,148816\tablefootmark{a}  &  $5840$  &  $4.14$  &  $-0.70$  &  $1.40$  &  $0.26$  &  $-0.11$  &  $-0.16$  &  $-0.12$  &  $-0.13 \pm 0.03$  &high-\alp&  L  \\
                &  $    $  &  $    $  &  $     $  &  $    $  &  $    $  &  $-0.03$  &  $-0.12$  &  $-0.05$  &  $-0.07 \pm 0.05$  &      &  N  \\
  HD\,159482  &  $5737$  &  $4.31$  &  $-0.73$  &  $1.30$  &  $0.30$  &  $-0.10$  &  $-0.11$  &  $-0.10$  &  $-0.10 \pm 0.01$  &high-\alp&  L  \\
                &  $    $  &  $    $  &  $     $  &  $    $  &  $    $  &  $-0.03$  &  $-0.08$  &  $-0.04$  &  $-0.05 \pm 0.03$  &      &  N  \\
  HD\,160693  &  $5714$  &  $4.27$  &  $-0.49$  &  $1.10$  &  $0.19$  &  $-0.09$  &  $-0.06$  &  $-0.14$  &  $-0.10 \pm 0.04$  &high-\alp&  L  \\
                &  $    $  &  $    $  &  $     $  &  $    $  &  $    $  &  $-0.04$  &  $-0.04$  &  $-0.10$  &  $-0.06 \pm 0.03$  &      &  N  \\

\end{tabular}
\end{table*}

\begin{table*}
\centering
Table \ref{tab2} continued.
\setlength{\tabcolsep}{0.2cm}
\begin{tabular}{lccccrrrrrcc}
\noalign{\smallskip}
\hline\hline
\noalign{\smallskip}
\noalign{\smallskip}
Star & \teff\ (K)  & \logg  & \feh  & $\xi$ & \alpfe & $5105$\AA  & $5218$\AA  & $5782$\AA  & \cufe  & Class & LFS\\
\noalign{\smallskip}
\hline
\noalign{\smallskip}
    HD\,177095  &  $5349$  &  $4.39$  &  $-0.74$  &  $0.90$  &  $0.31$  &  $-0.09$  &  $-0.10$  &  $-0.16$  &  $-0.12 \pm 0.04$  &high-\alp&  L  \\
                &  $    $  &  $    $  &  $     $  &  $    $  &  $    $  &  $-0.05$  &  $-0.09$  &  $-0.13$  &  $-0.09 \pm 0.04$  &      &  N  \\
    HD\,179626\tablefootmark{a}  &  $5855$  &  $4.19$  &  $-1.00$  &  $1.40$  &  $0.32$  &  $-0.33$  &  $-0.27$  &  $-0.22$  &  $-0.27 \pm 0.06$  &high-\alp&  L  \\
                &  $    $  &  $    $  &  $     $  &  $    $  &  $    $  &  $-0.22$  &  $-0.21$  &  $-0.12$  &  $-0.18 \pm 0.06$  &      &  N  \\
    HD\,189558\tablefootmark{a}  &  $5623$  &  $3.81$  &  $-1.12$  &  $1.40$  &  $0.35$  &  $-0.44$  &  $-0.40$  &  $-0.37$  &  $-0.40 \pm 0.04$  &  TD  &  L  \\
                &  $    $  &  $    $  &  $     $  &  $    $  &  $    $  &  $-0.32$  &  $-0.34$  &  $-0.25$  &  $-0.30 \pm 0.05$  &      &  N  \\
    HD\,193901\tablefootmark{a}  &  $5676$  &  $4.41$  &  $-1.07$  &  $1.40$  &  $0.17$  &  $-0.65$  &  $-0.60$  &  $-0.54$  &  $-0.60 \pm 0.06$  & low-\alp&  L  \\
                &  $    $  &  $    $  &  $     $  &  $    $  &  $    $  &  $-0.57$  &  $-0.57$  &  $-0.47$  &  $-0.54 \pm 0.06$  &      &  N  \\
    HD\,194598\tablefootmark{a}  &  $5926$  &  $4.32$  &  $-1.08$  &  $1.40$  &  $0.20$  &  $-0.43$  &  $-0.44$  &  $  ...$  &  $-0.44 \pm 0.01$  & low-\alp&  L  \\
                &  $    $  &  $    $  &  $     $  &  $    $  &  $    $  &  $-0.32$  &  $-0.38$  &  $  ...$  &  $-0.35 \pm 0.04$  &      &  N  \\
    HD\,230409  &  $5318$  &  $4.54$  &  $-0.85$  &  $1.10$  &  $0.27$  &  $ 0.00$  &  $-0.02$  &  $-0.07$  &  $-0.03 \pm 0.04$  &high-\alp&  L  \\
                &  $    $  &  $    $  &  $     $  &  $    $  &  $    $  &  $ 0.04$  &  $-0.01$  &  $-0.04$  &  $-0.00 \pm 0.04$  &      &  N  \\
    HD\,237822  &  $5603$  &  $4.33$  &  $-0.45$  &  $1.10$  &  $0.29$  &  $ 0.02$  &  $ 0.02$  &  $-0.02$  &  $ 0.01 \pm 0.02$  &high-\alp&  L  \\
                &  $    $  &  $    $  &  $     $  &  $    $  &  $    $  &  $ 0.06$  &  $ 0.03$  &  $ 0.02$  &  $ 0.04 \pm 0.02$  &      &  N  \\
  HD\,250792A  &  $5489$  &  $4.47$  &  $-1.01$  &  $1.10$  &  $0.24$  &  $-0.54$  &  $-0.48$  &  $-0.51$  &  $-0.51 \pm 0.03$  & low-\alp&  L  \\
                &  $    $  &  $    $  &  $     $  &  $    $  &  $    $  &  $-0.49$  &  $-0.47$  &  $-0.46$  &  $-0.47 \pm 0.02$  &      &  N  \\
\noalign{\smallskip}
\hline
\end{tabular}
\tablefoot{
Table \ref{tab2} is in the same format as Table \ref{tab1}. The stars without available copper abundance are not listed in the table.\\
\tablefoottext{a}{The stars that also have VLT/UVES spectra.}
}
\end{table*}

\end{document}